# Hegemonic structure of basic, clinical and patented knowledge on Ebola research: a US army reductionist initiative

David Fajardo-Ortiz[1*], José Ortega-Sánchez-de-Tagle[2] and Victor M Castaño[3*]

## Abstract

**Background:** Ebola hemorrhagic fever (Ebola) is still a highly lethal infectious disease long affecting mainly neglected populations in sub-Saharan Africa. Moreover, this disease is now considered a potential worldwide threat. In this paper, we present an approach to understand how the basic, clinical and patent knowledge on Ebola is organized and intercommunicated and what leading factor could be shaping the evolution of the knowledge translation process for this disease.

**Methodology:** A combination of citation network analysis; analysis of Medical heading Subject (MeSH) and Gene Ontology (GO) terms, and quantitative content analysis for patents and scientific literature, aimed to map the organization of Ebola research was carried out.

**Results:** We found six putative research fronts (i.e. clusters of high interconnected papers). Three research fronts are basic research on Ebola virus structural proteins: glycoprotein, VP40 and VP35, respectively. There is a fourth research front of basic research papers on pathogenesis, which is the organizing hub of Ebola research. A fifth research front is pre-clinical research focused on vaccines and glycoproteins. Finally, a clinical-epidemiology research front related to the disease outbreaks was identified. The network structure of patent families shows that the dominant design is the use of Ebola virus proteins as targets of vaccines and other immunological treatments. Therefore, patents network organization resembles the organization of the scientific literature. Specifically, the knowledge on Ebola would flow from higher (clinical-epidemiology) to intermediated (cellular-tissular pathogenesis) to lower (molecular interactions) levels of organization.

**Conclusion:** Our results suggest a strong reductionist approach for Ebola research probably influenced by the lethality of the disease. On the other hand, the ownership profile of the patent families network and the main researches relationship with the United State Army suggest a strong involvement of this military institution in Ebola research.

## Background

Ebola hemorrhagic fever (Ebola) is an acute viral disease with high lethality rates [1], which overwhelmingly affects populations in Sub-Saharan Africa [2]. The current Ebola disease outbreak is the most devastating one of the recorded History. It has killed more people that all the previous outbreaks combined [3].

Interestingly, Ebola is not considered a neglected tropical disease, but it is rather classified as an acute and emergent disease [4]. However, Ebola does shares three key features with other diseases recognized as neglected: first, this disease affects neglected populations, with no access to a proper health care [5], as the current outbreak evidences. It has been reported that the "clinical course of infection and the transmissibility of the virus are similar to those in previous EVD outbreaks" [3].

* Correspondence: davguifaj@yahoo.com; meneses@unam.mx
[1]Graduate program in Medical Sciences and Health, Universidad Nacional Autónoma de México, Mexico City, Mexico
[3]Centro de Fisica Aplicada y Tecnologia Avanzada, Universidad Nacional Autonoma de Mexico, Queretaro, Mexico
Full list of author information is available at the end of the article

Interestingly, the Ebola virus is reported to be genetically stable in the wild [6]. Therefore, the catastrophic dimension of the current outbreak could be mainly explained by the impoverishing of the national health systems [5]. It has even been suggested that some punctual control measures and a higher quality health system could have avoided many deaths [3,5]. Second, in spite of the first outbreak in 1979 [7], nowadays there are not approved vaccines or drugs to help the affected population [8]. This lack of approved vaccines and drugs evidences that innovation on Ebola has not been considered an attractive enough business for the pharmaceutical industry [9,10]. Finally, the third characteristic is that some authors have suggested a racial stigmatization against Sub-Saharan African-ancestry people as potential vectors of Ebola [11-13]. The possible stigmatization of Africa as an exporter of dangerous diseases like Ebola, is exemplified [11] by Richard Preston's non-fiction Bestseller "The Hot Zone" [14]. In this regard, Haynes states that: "Preston exploits post-Cold War insecurities about African contamination in the narrative structure of The Hot Zone. By employing a long established discourse about Africa as the "white man's grave", he inscribes these rare filoviruses as a genuine biological threat to the people and security of the United States" [11]. Meanwhile, Murdocca states on the case of a patient wrongly diagnosed with Ebola in Canada in 2001 that "representation of immigrants as vectors of disease is a useful and coercive tool in the project of justifying immigration reform and the state control of racial bodies" [12].

Besides these three aggravating features of Ebola, this disease is frequently considered a threat to the US national security [15,16]. Indeed, as an example, a US National Intelligence Council document, published in 2000, identified Ebola as a global threat [17]. In fact, some authors have even suggested the potential weaponization of the Ebola virus [18,19]. Moreover, recently, the US President, Barack Obama, stated in a letter addressed to the House of Representatives (Emergency Appropriations Request for Ebola for Fiscal Year 2015) the following:

> "The request also includes resources to strengthen global health security by reducing risks to Americans by enhancing the capacity of vulnerable countries to prevent disease outbreaks, detect them early, and swiftly respond before they become epidemics that threaten our national security" [20].

The neglected disease-like features and its condition as a threat to the US security would be the main elements of the socio-political context of Ebola. We hypothesized whether or not that these two elements could be influencing the way we understand and research Ebola. In this general context, this present piece of research has two main objectives. The first is to explore how -for the first time- the knowledge on Ebola is organized through the literature and patents networks. The second, we have investigated some indications of whether Ebola research is affected or not by the two above elements.

To have a better perspective on how Ebola research is organized, we have separated this exploration into three more specific questions. First, which are the main research fronts for Ebola and how they are intercommunicated. Research fronts are clusters of highly interconnected papers [21] putatively related to hidden colleges, i.e., virtual communities of researchers who cited each other and work on similar topics and share a similar way to research and understand the problems. Second, how the basic, translational and clinical knowledge on Ebola is structured. Finally, what could be the dominant design for Ebola according the structure of the patent families networks. A dominant design is a system of paradigmatic and standardized components or features within a particular product class [22]. For example, pegylated liposomal doxorubicin, a liposomal old drug, would be the dominant design for cancer nanotechnologies [23]. Because there are no approved pharmaceutical treatments for Ebola yet, we decided to consider as a dominant design the most prevalent technological option that simultaneously appears in the literature and patent networks.

## Methods

Our research is only based on the analysis of publicly available secondary information: Abstracts of papers (all of these are available in the Medical Literature Analysis and Retrieval System Online MEDLINE© of the United States National Library of Medicine) and the content of patents (public documents which are available in The United States Patent and Trademark Office (USPTO), European Patent Office (EPO), the World Intellectual Property Organization (WIPO) and other national patent offices). Therefore, this research did not require the approval of an ethics committee. We have previously developed a combination of methodologies to explore the network structure of scientific literature, which allows identify the main research fronts of a certain field and how these are interconnected each to other [23,24]. Also, these methodologies can map the knowledge translation process through the literature networks [23,24]. In a previous research on liposomes for cancer therapy, we adapted these methodologies to analyze patents networks [25]. The following steps of the above methodologies were followed in this study:

1. A search of research papers on Ebola was performed in the Web of Science (WOS) [26] during October, 2014. The search criteria were the following: Title:

　　Ebola; Document types: Article; Timespan: All years; Indexes: SCI-EXPANDED, SSCI, A&HCI, CPCI-S, CPCI-SSH, BKCI-S, BKCI-SSH. We founded 752 papers.
2. We selected the 20% most cited papers (151). These 151 documents accumulated the 63% (18,260 of 28,970) of the citations that the 752 papers on Ebola have received. It is important to point out that citations in science tend to be distributed according to a Zipf's law [27]. That is, a very small number of papers receive a large quantity of citations, whereas the most of papers have few citations or none. We selected these 151 papers on Ebola because they are a small, workable and readable number of papers that accumulate most of the communication process through the citations network.
3. The software Histcite [28] was used to build a citation network model of the selected papers. Cytoscape [29], an open source software, was used as a platform for visualization and analysis of the network model. Clust&See [30], a Cytoscape plug-in, was used to divide the network model in clusters (according to the Newman modularity that defines clusters as "groups of vertices within which connections are dense but between which they are sparse" [31]), which are putatively related to different research fronts.
4. The selected papers were searched in GOPubmed [32]. This search engine semantically analyzed the papers (title and abstract) and labeled them with Gene Ontology (GO) [33] and Medical Subject Heading (MeSH) [34] terms. We a priori defined as clinical terms all MeSH terms that belong to the next higher hierarchy categories: "Diagnosis", "Therapeutics", "Surgical Procedures, Operative", "Named Groups", "Persons" and "Health Care". We calculated the rate of clinical terms vs. non-clinical for every paper in the citation network model. The MeSH and GO terms that were common for most of the papers of each cluster were identified. The most characteristic of these GO and MeSH terms were used to label and differentiate the clusters.
5. The modes (papers) were colored according to a color (from red to blue), which is a function of the clinical terms rate. The network model was displayed using the "spring embedded" algorithm.
6. The abstracts of the papers in the network model were analyzed using KH Coder, a software for quantitative content analysis [35-37]. KH Coder was used to perform two different analyses in order to compare the content of the clusters. First, the top 10 distinctive world of each cluster were identified using the Jaccard index as a distinctiveness measure. Second, a correspondence analysis was performed. "Correspondence analysis is a descriptive/exploratory technique that uses a simple two-way and multi-way contingency table" [37].
7. A search of patent families (i.e. a set of patents that refers to the same invention) were performed in the Derwent Innovations Index [38] during October, 2014. The search criteria were the following: TITLE: Ebola; Timespan: All years; Indexes: CDerwent, Ederwent and MDerwent. We founded 102 patents. We selected the 20% most cited of the patent families (21) which received 77.6% of the citations (146 of 188). Note that we are just considering citations among patent families without citations from other sources, like research papers.
8. We build a network model of inter-citation for the selected patent families. The model was visualized using Cytoscape. AllegroMCODE [36] was used to identify highly interconnected (dense) regions of the network model. These dense regions are related to types of inventions that could share the same inventors or assignees.
9. The title and abstract of the patent families in the network model were analyzed using the semantic annotator of GOPubmed. The main MeSH and GO terms associated with the network model and dense regions were identified.

## Results

We built two different semantically analyzed network models (maps). The first map displays the general structure and inter-communication of basic, translational and clinical research on Ebola, while the second map shows how the patented knowledge is structured and what could be the dominant design for anti-Ebola therapy. These maps are separately described in what follows.

### Ebola research map

A citation network of 150 top cited papers on Ebola was built (Figure 1). Clust&See divided the network in seven clusters (Figure 1). These clusters of papers are mainly organized around three structural proteins of the Ebola virus, according to the GO and MeSH terms distribution (Figure 1 and Table 1). Also, there are a cluster of epidemiological-clinical papers around the term "disease outbreak" and one cluster of papers on pathogenesis. The clusters are numbered according to their size rank and named with their most representative GO or MeSH terms.

　　Cluster 1 consists of 33 papers and 173 inter-citations. These papers exhibited an average rate of clinical terms of 1.4%. The GOPubmed terms that distinguish this cluster are "glycoproteins" and "viral envelope" (Table 1). 14 of these papers were published in the Journal of Virology. The paper with the highest in-degree (the most cited by the papers in the cluster) is "The virion glycoproteins of Ebola viruses are encoded in two reading frames and are expressed through transcriptional editing" [6]. Cluster 1 papers clearly form a basic research front organized around the Ebola virus structural glycoprotein.

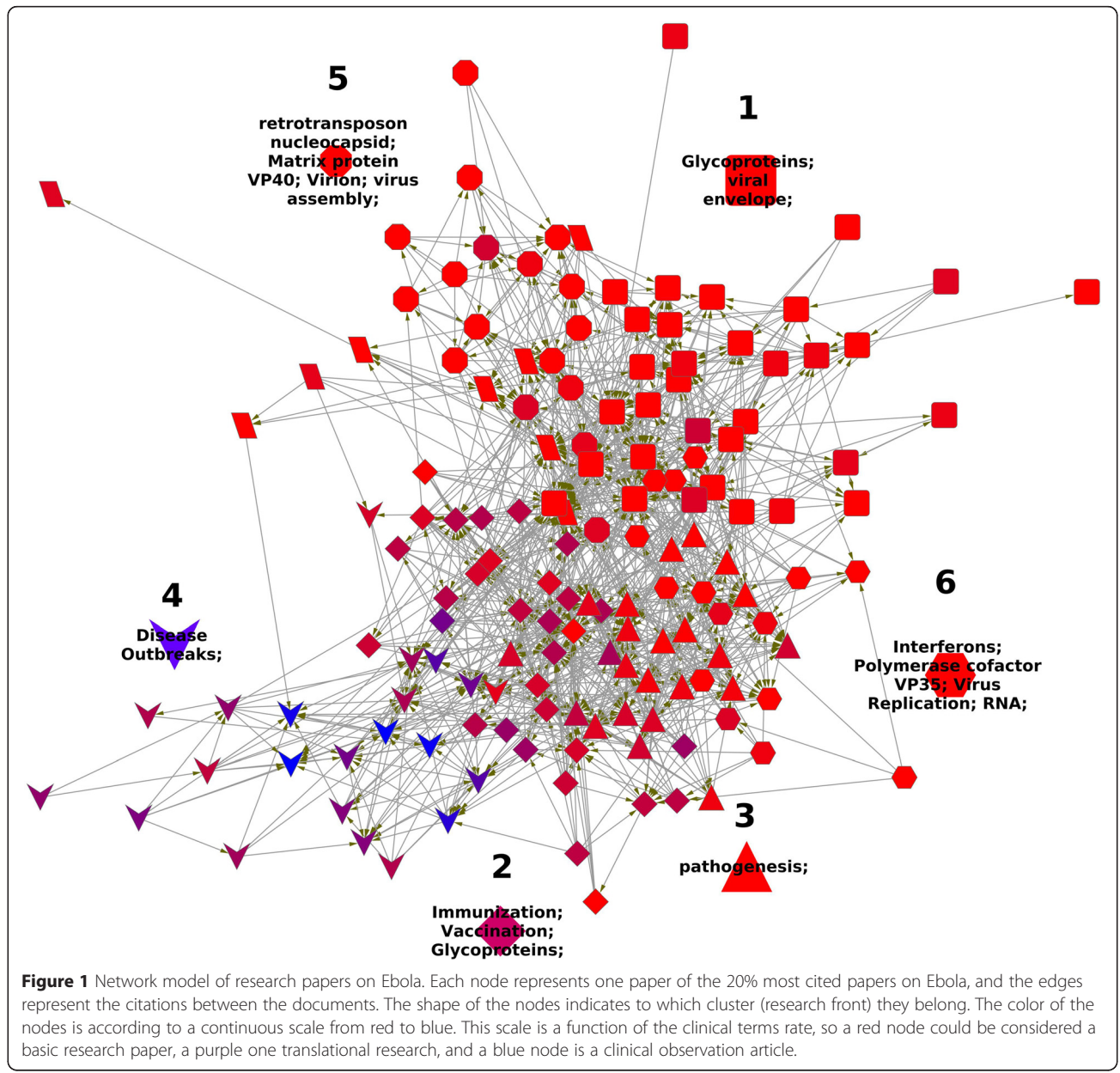

**Figure 1** Network model of research papers on Ebola. Each node represents one paper of the 20% most cited papers on Ebola, and the edges represent the citations between the documents. The shape of the nodes indicates to which cluster (research front) they belong. The color of the nodes is according to a continuous scale from red to blue. This scale is a function of the clinical terms rate, so a red node could be considered a basic research paper, a purple one translational research, and a blue node is a clinical observation article.

Cluster 2 consists of 31 papers and 147 inter-citations. These papers exhibited an average rate of clinical terms of 11.8%. Its distinguishing GOPubmed terms are "immunization", "vaccination", "vaccines" and "glycoproteins" (Table 1). Peter Jahrling, currently Chief Scientist at The USA National Institute of Allergy and Infectious Diseases (NIAID), is author of 14 of these papers. The papers with the highest in-degree are "A Mouse Model for Evaluation of Prophylaxis and Therapy of Ebola Hemorrhagic Fever" [39] and "Immunization for Ebola virus infection" [40]. The latter paper reported an research about the Ebola nucleoproteins and glycoproteins ability to protect against infection. Cluster 2 papers are mainly pre-clinical research focused on the discovery of potential immunotherapies based in glycoproteins and other molecular entities.

Cluster 3 consists of 23 papers and 117 inter-citations. The average rate of clinical terms is 4.4%. "Pathogenesis" is the distinguishing GOPubmed term (Table 1). Thomas Geisbert, Professor, Microbiology & Immunology, is coauthor of 9 papers belonging to this cluster. It is important to note that his laboratory "focuses on the pathogenesis of emerging and re-emerging viruses that require Biosafety level (BSL)-4 containment and on the development of countermeasures against these viruses" [41]. The paper with the highest in-degree is "Association of Ebola-related Reston virus particles and antigen with tissue lesions of monkeys imported to the United States" [42]. This cluster consists of basic research papers mainly focused on pathogenesis.

Table 1 Main GO and MeSH terms of each cluster

| Main GO and MeSH terms | Number of papers |
|---|---|
| Cluster 1 | |
| Ebolavirus | 33 |
|     Viruses | 32 |
|     Glycoproteins | 31 |
|     Humans | 26 |
|     Viral envelope | 24 |
|     Animals | 23 |
|     Cell Line | 18 |
| Cluster 2 | |
| Ebolavirus | 30 |
|     Animals | 30 |
|     Viruses | 30 |
|     Hemorrhagic Fever, Ebola | 28 |
|     Immunization | 20 |
|     Vaccination | 20 |
|     Vaccines | 20 |
|     Immunity | 18 |
|     Mice | 18 |
|     Glycoproteins | 18 |
|     Antibodies, Viral | 15 |
|     Humans | 15 |
|     Hemorrhage | 15 |
| Cluster 3 | |
|     Viruses | 19 |
|     Ebolavirus | 19 |
|     Humans | 13 |
|     Animals | 13 |
|     Hemorrhagic Fever, Ebola | 12 |
|     Pathogenesis | 12 |
| Cluster 4 | |
| Ebolavirus | 14 |
|     Humans | 14 |
|     Viruses | 13 |
|     Hemorrhagic Fever, Ebola | 12 |
|     Disease Outbreaks | 11 |
| Cluster 5 | |
| Viruses | 14 |
|     Ebolavirus | 13 |
|     Proteins | 13 |
|     Humans | 10 |
|     Viral Proteins | 9 |
|     Retrotransposon nucleocapsid | 9 |
|     Matrix protein VP40 | 9 |
|     Virion | 9 |
|     Virus assembly | 9 |
| Cluster 6 | |
| Viruses | 13 |
|     Ebolavirus | 12 |
|     Humans | 10 |
|     Viral Proteins | 9 |
|     Genes | 9 |
|     Proteins | 9 |
|     Animals | 9 |
|     Interferons | 8 |
|     Polymerase cofactor VP35 | 8 |
|     Virus Replication | 8 |
|     RNA | 8 |

Cluster 4 consists of 22 papers and 76 inter-citations. The average rate of clinical terms is 27.8%. "Disease outbreaks" is the distinguishing GOPubmed term (Table 1). The paper with the highest in-degree is "Isolation and partial characterization of a new strain of Ebola virus" [43]. It is important to note that this paper reported "the first time that a human infection has been connected to naturally-infected monkeys in Africa" [40]. Cluster 4 is clinical-epidemiological set of papers focused on Ebola disease outbreaks. Thomas Ksiazek, Director of high containment laboratory operations for the Galveston National Laboratory at the University of Texas, is the main author of this cluster. He is the author of 9 papers in cluster 4.

Cluster 5 consists of 17 papers and 66 inter-citations. The average rate of clinical terms is 1.9%. Its distinguishing GOPubmed terms are "retrotransposon nucleocapsid", "matrix protein VP40", "vrion" and "virus assembly" (Table 1). The papers with the highest in-degree are "Vesicular release of Ebola virus matrix protein VP40" [44] and "Ebola virus VP40-induced particle formation and association with the lipid bilayer" [45]. Cluster 4 is basic research focused in the Ebola virus structural protein VP40.

Cluster 6 consists of 15 papers and 51 inter-citations. The average rate of clinical terms is 1.2%. Its distinguishing

GOPubmed terms are "Interferons", "polymerase cofactor VP35" and "virus replication". The papers with the highest in-degree are "Comparison of the transcription and replication strategies of Marburg virus and Ebola virus by using artificial replication systems" and "The Ebola virus VP35 protein functions as a type IIFN antagonist" [46]. Cluster 4 is basic research focused in the Ebola virus structural protein VP35.

Cluster 7 consists of 9 papers and 19 inter-citations. The average rate of clinical terms is 1.2%. The Gopubmed terms are shared with the rest of clusters, i.e., they are too general to consider cluster 7 related to a putative research front.

Figure 2 summarizes how these cluster are organized together. Each node represent one cluster and the arrows are formed by the sum of the inter-citations between two clusters. In order to make it more readable, we hid the links formed with less than 30 inter-citations. Cluster 3 (pathology) is the hub of the Ebola research map. The strongest connection is between cluster 3 and cluster 2 (vaccines and glycoproteins). The second strongest connection is between cluster 1 (glycoproteins and viral envelope) and cluster 3. The basic and pre-clinical researchs on glycoproteins (clusters 1 and 2) are connected by the third biggest inter-citation. Clusters 6 (VP35) and 4 (disease outbreaks) are mainly connected to cluster 3. Cluster 4 (VP40) is mostly connected to cluster 1. Cluster 7 is just an appendix of cluster 1. The citations among the clusters show a directionality from cluster 4 "disease outbreaks" to cluster 3 "pathogenesis" to the clusters that are related to Ebola virus proteins.

110 papers of the network model were coauthored by researchers working in The United States of America.

## Qualitative content analysis of the clusters

The qualitative content analysis is consistent with that observed in the GO and MeSH term distribution (Table 1). Cluster 4 most distinctive words are "outbreak", "patient", "case" and "human" (Table 2). The correspondence analysis plot showed that cluster 3 has the most similar content to cluster 4 (Figure 3). This matches with the fact that cluster 4 is mainly connected to the rest of the citation network through cluster 3 (Figure 2). The GO and MeSH terms and the quantitative content analysis suggested that cluster 4 is related to a clinical-epidemiological research front on Ebola. Cluster 3 most distinctive words are "cell", "infection", "level", "response", "viral" and "endothelial" (Table 2). The correspondence analysis placed cluster 3 in a intermediate position between cluster 1 and cluster 4 (Figure 3), which resembles the inter-citation structure among the clusters (Figure 2). The quantitative content analysis and the GO and MeSH terms distribution suggest that cluster 3 is basic research focused on the host immune system-virus pathogenic interaction. Clusters 1, 5 and 6 most distinctive words are related to molecular-level concepts like "Glycoprotein", "virus", "VP40" (Viral protein 40), "particle", "motif", "VP35" (Viral protein 35) or "interferon" (Table 2). These three clusters are practically equidistant in the concurrence analysis plot (Figure 3). According to the MeSH and GO terms distribution and the quantitative content analysis, these three clusters are basic research focused on the interaction of the viral proteins and the molecular machinery of the host immune cells. Finally, cluster 2 most distinctive words are "vaccine", "ebola", "virus", "challenge", "protect" and "lethal" (Table 2). Cluster 2 is located far away from the other ones in the correspondence analysis plot, i.e., its content is quite different from the rest of clusters. The distribution of MeSH and GO terms (Table 1) and the qualitative content analysis (Table 2 and Figure 3) suggest that cluster 2 is translational research focused in the development of vaccines.

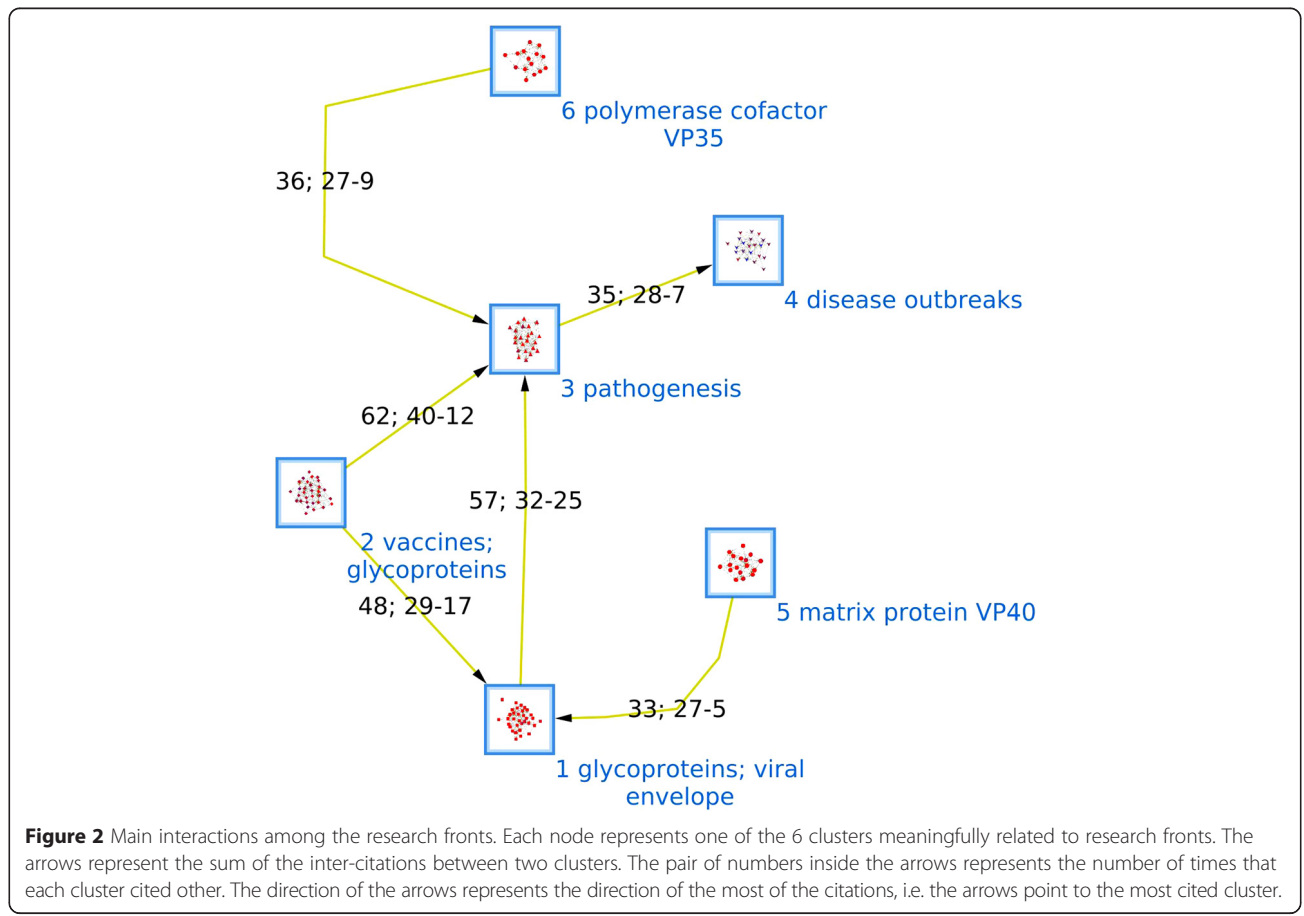

Figure 2 Main interactions among the research fronts. Each node represents one of the 6 clusters meaningfully related to research fronts. The arrows represent the sum of the inter-citations between two clusters. The pair of numbers inside the arrows represents the number of times that each cluster cited other. The direction of the arrows represents the direction of the most of the citations, i.e. the arrows point to the most cited cluster.

Table 2 List of the top 10 distinctive world of each cluster

| Cluster 1 | Distinctiveness | Cluster 2 | Distinctiveness | Cluster 3 | Distinctiveness | Cluster 4 | Distinctiveness |
|---|---|---|---|---|---|---|---|
| GP | .277 | Vaccine | .273 | Cell | .153 | Outbreak | .216 |
| Cell | .222 | Ebola | .210 | Infection | .131 | Patient | .186 |
| Virus | .201 | Virus | .202 | Level | .102 | Case | .164 |
| Glycoprotein | .177 | Challenge | .163 | Response | .099 | Human | .096 |
| Ebola | .167 | Protect | .154 | Viral | .098 | Antigen | .093 |
| Entry | .130 | Lethal | .143 | Endothelial | .084 | Hemorrhagic | .091 |
| Viral | .088 | Infection | .134 | Replication | .079 | Fever | .090 |
| Fusion | .086 | Mouse | .122 | Patient | .077 | Assay | .090 |
| GP2 | .081 | Antibody | .108 | Study | .076 | Serum | .087 |
| Surface | .072 | Animal | .102 | Macrophage | .076 | Antibody | .087 |
| **Cluster 5** | **Distinctiveness** | **Cluster 6** | **Distinctiveness** | | | | |
| VP40 | .406 | Protein | .144 | | | | |
| Bud | .254 | EBOV | .136 | | | | |
| Protein | .200 | VP35 | .133 | | | | |
| Particle | .156 | Gene | .094 | | | | |
| Motif | .143 | RNA | .093 | | | | |
| Membrane | .126 | Interferon | .087 | | | | |
| Assembly | .121 | VP24 | .083 | | | | |
| Ebola | .116 | IFN | .082 | | | | |
| Matrix | .106 | Response | .079 | | | | |
| vlp | .106 | Activation | .078 | | | | |

### Ebola patenting map

16 of the selected patent families form four small and simple networks (Figure 4). AllegroMCODE identified just one densely connected region formed by four patent families (Figure 4, yellow nodes). The United States Secretary of the Army is the leading assignee in this central region (Figure 4). A more recent search (December 2014) in the Derwent Innovations Index show that patent families in the densely connected region are the first (Derwent Primary Accession Number, DPAN: 2000–160677), second (DPAN: 1999–405117) and fourth (DPAN: 2004–226835) most cited. The semantic analysis showed these four inventions are related to vaccine development, antibodies and viral DNA (Table 3). Basic Gopubmed terms like "Infection", "Cells" and "viral reproduction" are commons to the most of the 16 interconnected patent families (Table 3).

## Discussion

### Elements for the interpretation of results

We consider that there are two necessary conceptual elements to interpret our results. First of all, we searched for papers and patent families with the world "Ebola" in their titles. We did not a priori search for papers on pathogenesis of Ebola disease, Ebola virus genome or health policy research on Ebola, we searched just for "Ebola". Second, we focused in the most cited documents. Scientific communication has a hegemonic structure reflexed in the power law distribution of its citations [27]. Hegemony in science means that there are some ideas or concepts much more influential than others. The most cited papers citing each other constitute the paradigmatic and hegemonic body of a particular field of study. Similarly, the most cited inventions trend to show what the dominant design is. An additional element is that certain concepts are related to particular research fronts. Therefore, in this research we looked for the ideas, research fronts and technological strategies that form the paradigmatic view of Ebola. We discuss below how this paradigm is organized and how the nature and context of Ebola (as a acute disease affecting neglected populations and a national-global security threat) partially explains its knowledge structure.

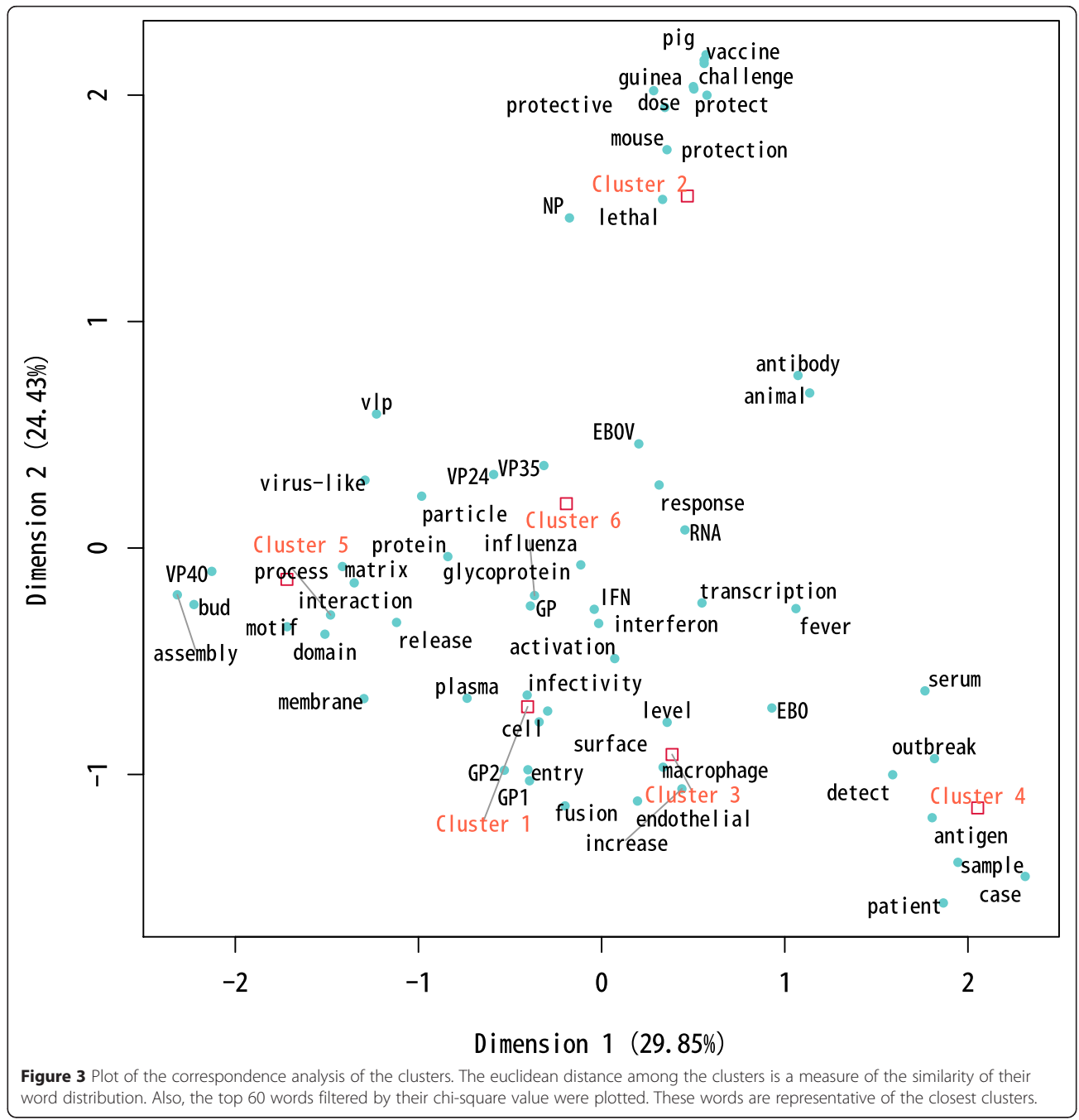

**Figure 3** Plot of the correspondence analysis of the clusters. The euclidean distance among the clusters is a measure of the similarity of their word distribution. Also, the top 60 words filtered by their chi-square value were plotted. These words are representative of the closest clusters.

### The big picture of Ebola research

Figure 2 illustrates how this paradigmatic view is structured. We identified two relevant features in the literature network model. Firstly, the reductionist directionality of the biggest interactions among the putative research fronts, i.e., the knowledge seems to flow from "disease outbreaks" (A population-to-individual level research front) to "pathogenesis" (A cellular level research front) to the viral protein-related research fronts (biomolecular level). Secondly, the only one research front that is related to the development of potential therapeutic alternatives (cluster 2 "vaccines") is at a cellular-biomolecular level and this research front strongly cites papers belonging to the "glycoprotein" and "pathogenesis" research fronts. These two features are discussed in detail next.

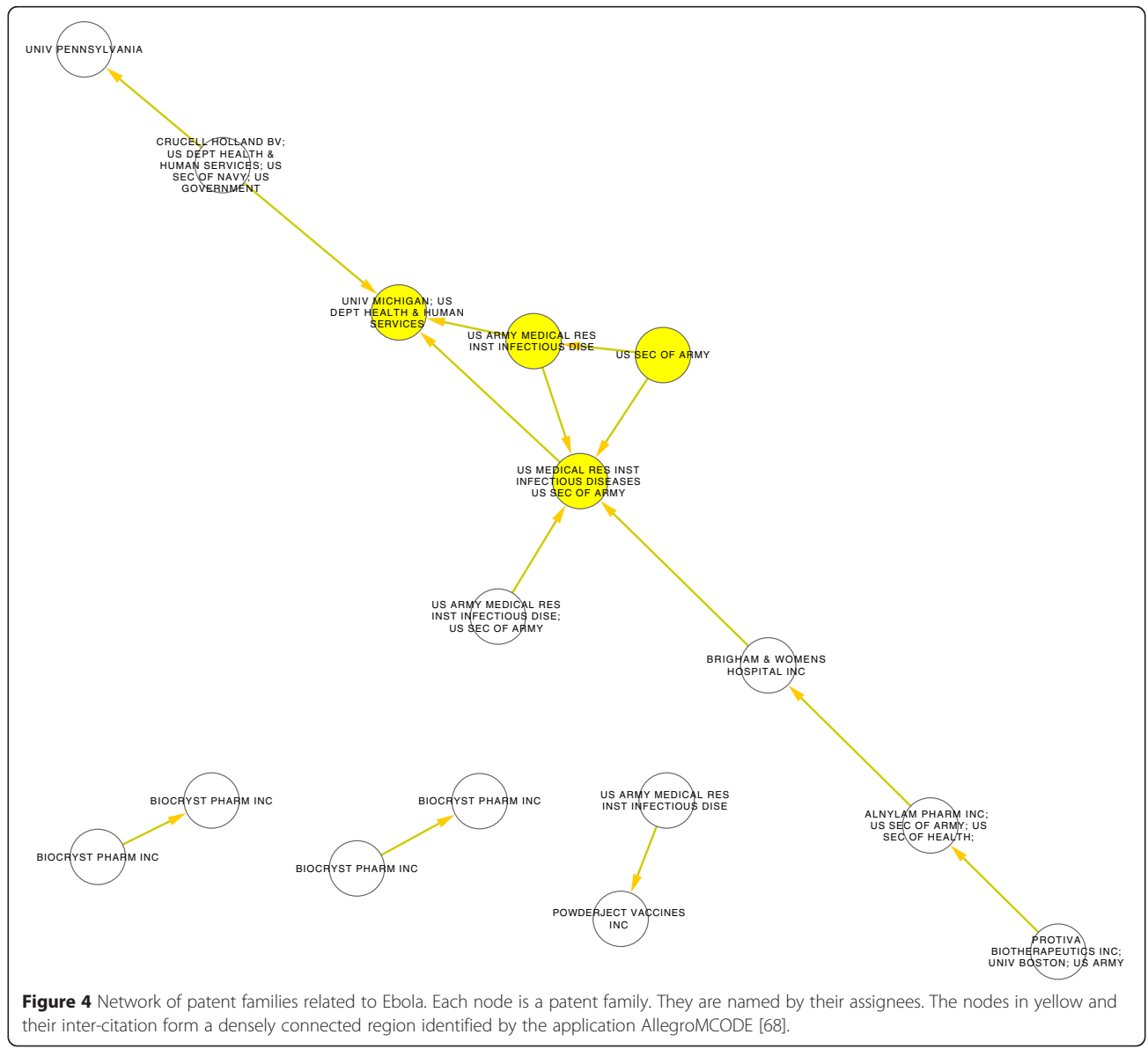

**Figure 4** Network of patent families related to Ebola. Each node is a patent family. They are named by their assignees. The nodes in yellow and their inter-citation form a densely connected region identified by the application AllegroMCODE [68].

### The reductionist directionality and the contextualization

Apparently, there is a reductionist directionality among the research fronts from higher to lower organization levels (Figure 2). The exception could be the interaction between clusters 3 and 1, which is mostly bidirectional. Reduction means that a more complex phenomenon is explained by the interaction of other phenomena in a simpler or more fundamental level. Thus, "Disease outbreaks" research (Cluster 4) is, in a way, the phenomenological description of the disease that should be explained by the more fundamental "pathogenesis" research (Cluster 3). In turn, the "pathogenesis" phenomenon would be explained by the research on the structural Ebola virus proteins GP, VP40 and VP35 (Clusters 1, 5 and 6 respectively). On the other hand, "pathogenesis" research would be contextualized by the "disease outbreaks" research, and viral proteins research, in turn, would be contextualized by the "pathogenesis". In this research, contextualization means that the interactions occurring in a lower lever are correctly interpreted as long as they are considered as a part of a high level phenomenon. In order to clarify the idea of contextualization a passage of a paper written by Thomas W. Geisbert et al. is used as an example (this paper belongs to the cluster 3 "pathogenesis"):

> "Animal models that adequately reproduce human EBOV HF are clearly needed to gain further insight into the pathogenesis of this disease. [...] two rodent models of EBOV infection are not ideal for studying human EBOV HF; neither mice nor guinea pigs exhibit the hemorrhagic manifestations that characterize human EBOV infections" [47].

Table 3 Main GO and MeSH terms of the all connected patent families and the densely connected region

| Main GO and MeSH terms | Number of papers |
| --- | --- |
| All the connected patent families | |
| Viruses | 16 |
| Ebolavirus | 15 |
| Infection | 11 |
| Cell | 10 |
| Cells | 10 |
| Viral reproduction | 10 |
| Acids | 8 |
| Methods | 8 |
| RNA | 8 |
| Vaccination | 8 |
| Vaccines | 8 |
| The densely connected region | |
| Biosynthetic process | 4 |
| Ebolavirus | 4 |
| Vaccination | 4 |
| Vaccines | 4 |
| Viruses | 4 |
| Acids | 3 |
| Antibodies | 3 |
| Antibodies, Viral | 3 |
| Antigen binding | 3 |
| Cell | 3 |
| Cells | 3 |
| DNA | 3 |
| DNA, Viral | 3 |
| Host | 3 |
| Infection | 3 |
| Viral reproduction | 3 |

This paragraph clearly notes that the disease manifestation in the patient is the context of Ebola pathogenesis.

The discoveries at the cellular and molecular levels are relevant, useful and meaningful as long as animal models resemble the disease in the patients. So, because of the context, we could consider that cynomolgus macaque is a more adequate animal model for Ebola research than rodent models. On the other hand, the disruption of the behavior of some types of leukocytes provoked by Ebola virus is a key component of the explanation of Ebola disease.

This kind of "reduction-contextualization" relationship could be also observed between "pathogenesis" and the viral proteins research fronts. A example of this type of relation is noted in a paper of cluster 6 "polymerase cofactor VP35" citing another document of cluster 3 "pathogenesis".

Mahanty et al. state in a paper belonging to the "pathogenesis" the next conclusion (Please note that DC is the abbreviation of dendritic cells in the passage):

> "Ebola and Lassa viruses infect human monocyte-derived DC and impair their function [...] These data represent the first evidence for a mechanism by which Ebola and Lassa viruses target DC to impair adaptive immunity" [48].

Basler et al. cited the previous paper in this paragraph:

> "Ebola virus infection has recently been reported to impair human dendritic cell function. It is possible that the ability of VP35 to inhibit IFN production may contribute to the ability of Ebola virus to inhibit dendritic cell functions" [49].

Dendritic cells impairment by Ebola virus is a key element of the explanation of Ebola disease. However, a reductionist approach requires a more fundamental (biomolecular) explanation: VP35 inhibition of anti-viral genes could be a factor that explains the dendritic impairment and therefore these could be fundamental to understand Ebola disease. Oppositely cellular behavior and communication are the context of research on VP35.

### From lab bench (glycoprotein) to bedside (vaccines)

According to our results, Ebola research does not have a applied clinical research front, unlike cancer research. However, it does have a pre-clinical or translational research front focused on vaccines development with virus glycoprotein as the main target: Cluster 2 "vaccines" (Figures 1, 2, 3 and Tables 1 and 2). The papers on vaccines strongly cite "pathogenesis" and "glycoproteins" documents (Figure 2). These results suggest that the development of medical interventions for Ebola disease are only considering the knowledge at the cellular-molecular level, at least in terms of organized research fronts and the communication among these. That is, the reductionist approach is dominating the knowledge translation process (NT). It is important to note that we operatively define NT as the communication between basic (discovery) research and applied clinical research through the inter-citation [23,24]. In the case of Ebola research, KT would be the inter-communication that the "vaccines" research front has with "pathogenesis" and "glycoproteins". A couple of examples of "vaccines" papers citing documents of the two basic research fronts are briefly analyzed below.

The first example shows how the knowledge on apotosis impacts on the development of vaccines. A "pathogenesis" paper reported that massive apoptosis happened in fatal human Ebola disease as follows:

> "We compared the immune responses of patients who died from Ebola virus disease with those who survived [...] DNA fragmentation in blood leukocytes and release of 41/7 nuclear matrix protein in plasma indicated that massive intravascular apoptosis proceeded relentlessly during the last 5 days of life" [50].

The above research is cited by a "vaccines" paper in order to argument that primate models are more suitable than rodent models for vaccine testing. Here the citing paragraph:

> "..rodent models of EBOV hemorrhagic fever do not consistently predict efficacy of candidate vaccines in nonhuman primates [...] Lymphocyte apoptosis was not reported to be a prominent feature of EBOV infection in mice or guinea pigs but was a consistent feature of disease in humans and nonhuman primates" [51].

These two previous paragraphs show how reductionism is influencing the KT process in Ebola research. That is, differences and similarities between humans and animal models at the cellular-molecular level, like the lymphocyte apoptosis, are what would predict the success or failure of a treatment.

An important part of the knowledge translation process (the interaction between clusters 1 and 2) is dominated by the research on one single biomolecule: the Ebola virus glycoprotein. The glycoprotein is simultaneously considered the key determinant of Ebola disease and the main component in the development of anti-Ebola vaccines. A cluster 2 of papers citing a cluster 1 document provides an example of the importance of this viral protein.

Yang et al. reported the importance of the glycoprotein mucin domain to explain the pathogenesis as follows:

> "...synthesis of the virion glycoprotein (GP) of Ebola virus Zaire induced cytotoxic effects in human endothelial cells in vitro and in vivo [...] These findings indicate that GP, through its mucin domain, is the viral determinant of Ebola pathogenicity and likely contributes to hemorrhage during infection" [52].

Martin et al. employed the above mentioned knowledge in order to design a safer glycoprotein-based vaccine through deletions in the pathogenic region of the viral protein. The citing paragraph is:

> "The Ebola virus GP genes expressed by plasmid DNA constructs in this vaccine contain deletions in the transmembrane region of GP that were intended to eliminate potential cellular toxicity observed in the in vitro experiments using plasmids expressing the full-length wild-type GPs" [53].

Finally, is important to notice that the reductionist biomolecular approach of Ebola research is necessarily coordinated with pharmaceutical strategies of intervention, like vaccines and other immunotherapies. According to our patents and scientific literature analyses, vaccines are the dominant design for Ebola, i.e., they are the hegemonic strategy to fight Ebola (see Figures 1, 2, 3 and Tables 1, 2, 3). Importantly, this strategy is fundamentally based on pre-clinical knowledge. In this research we did not find an organized body of clinical evidence (i.e., controlled trials, cohort studies or ecological studies). Instead, we found an organized body of mechanistic or pre-clinical evidence (rodent models and experiments involving non-human primates): the cluster 2. Previously, we had identified a full knowledge translation process for cervical cancer [54] and cancer nanotechnologies [23,24], i.e., basic research connected to clinical research through a translational research field. This is not the case with Ebola research, which has an incomplete knowledge translation process. This is important because the level of evidence is a key factor for decision-making in health [55]. Currently, there is a debate on the ethical use of experimental intervention on Ebola [56]. Some authors consider ethically valid the use of vaccines even though there is not clinical evidence to support it [57]. They consider that the dramatic dimension of the current outbreak makes the experimental pharmaceutical interventions necessary [57]. However, because these experimental interventions could imply unknown health risks [58] it is important to examine how the knowledge on Ebola is organized and who are the dominant stakeholders that influence its organization. Our results indicate that the US Army is the main assignee and the main research institution on Ebola. Particularly, the content of the patents (Table 3) and the leadership of US Army researchers in the clusters 2 and 3 suggest a possible connection between a national security paradigm and a highly sophisticated reductionist approach. In this regards, Colonel Erin P. Edgar commander of the United States Army Medical Research Institute of Infectious Diseases state the following:

> "...it is also clear that USAMRIID plays a critical role in the status of our country's preparedness for biological terrorism and biological warfare. While our primary mission is to protect the warfighter, our research benefits civilians as well" [59].

If these experimental intervention on Ebola were originally designed to satisfy the requirement of a national security agenda, how then we could be sure that these immunological technologies are the best option for the people of Sub-Saharan countries affected by the current outbreak?

### The reductionist approach and the lethal nature of Ebola disease

Unlike research on cancers and possibly other chronic diseases, it is difficult to study the Ebola disease as a condition of the patient (clinical knowledge). That is, in the case of Ebola there are not yet organized research fronts on treatment outcome (humans), survival, health policy or quality of life because of the acute and lethal nature of the disease. There is not enough time to focus on the patient suffering an acute disease unlike chronic diseases, which could last years. Ebola lethality could be enhancing a reductionist approach to study the disease mainly through two mechanisms. Firstly, biosafety level 4 requirement to work with Ebola virus could hamper some studies [60]. In order to avoid that problem researchers disassemble the Ebola virus to separately work with the structural proteins. For example, researchers uses recombinant vesicular stomatitis virus [60], replication-deficient adenovirus [61] or a plasmid containing genes of viral proteins [62]. Secondly, rodent models are the most feasible for high security biocontainment facilities [63] but they do not properly resemble the disease in human as non-human primates do [63,64]. Some mouse models are genetically modified in order to study specific molecular-cellular interaction among the virus and the host [63,64]. That is, the use of genetically-modified mouse model furthers the partition of the knowledge on the disease in molecular interaction and then synthetically rebuilds the pathogenesis (the whole) through the parts.

The reductionism approach of Ebola research may be lessened in the future due to two fundamental events: The emergence of approved and effective anti-Ebola drugs, and the current outbreak (2013–2015). First, the availability of pharmaceutical countermeasures could increase the chances of survival of the patients. In turn, higher survival rates would allow the emergence of research fields focused in the patient. Second, the dramatic differences between the current and the previous outbreaks can not be explained by biological phenomena but socioeconomic and environmental changes affecting the national health systems of the affected countries.

### Involvement of the US Army in Ebola Research & Development (R&D)

Two of our results suggest that Ebola R&D is influenced by the fact that Ebola disease is consider a national or global security threat, i.e., Ebola disease would be a military interest. The first result is that the main assignee in the patenting network is the United State Army (Figure 2). The second result is the overwhelming leadership of the United States in Ebola Research (110 of 150 top papers are authored by a researchers based in that country). The three most important authors of Ebola research namely, Peter Jahrling, Thomas Geisbert and Thomas Ksiazek (they are the leaders of clusters 2, 3 and 4, respectively) are or have been related to the United States Army Medical Research Institute of Infectious Diseases. It is clear the relevant participation of the US army in the Ebola R&D. But the important question is whether this implication of the Army is affecting the way Ebola is researched and understood. Our results suggest, at least, a possible relation between reductionism and the involvement of the US Army in Ebola research. The central and organizing patent family (DPAN: 2000–160677), which is owned by the United State Army, reported as an invention key epitopes and sequences of the structural viral proteins GP, NP, VP24, VP30, VP35 and VP40. The GO and MeSH terms related to the patents families (Table 3) that are mainly owned by the US army show that the inventions are conceptualized at a molecular-cellular level. Using Ebola virus proteins as a target of vaccines and other immunotherapies is the dominant design that could be being promoted by the military institution.

Ebola shares key features with neglected diseases. The social dimension of Ebola and the lethal nature of the disease could partially explain why there is not currently research focused in the patient and the observed reductionist and military orientation of Ebola research. Ebola R&D is not an attractive business to the pharmaceutical industry because its too expensive and there is not a large enough market [9,10]. However, Ebola is still considered a threat to the national security of United States ant therefore Ebola research is powered with public funds through the US army [65]. The US army, the main assignee, is currently looking for the development of drugs and biotechnological tools, a sort of magic bullets, to treat Ebola, which could be only possible through a reductionist approach.

### The gap between Ebola research and the multiple stakeholders expectancy

Previously, we have mentioned that the hegemonic knowledge on Ebola is mainly aimed to satisfy the requirement of a national security agenda. However, there is a plethora of stakeholders at different levels that could need different technological alternatives and approaches on Ebola: the affected communities, women, children, local governments, neighbor countries, the African Union, the Word Health Organization, the US government, etc. In this regard, Daniel Sarewitz and Roger A. Pielke Jr. conceptualized the relationship between the supply and demand of knowledge with two key questions in a 2x2 matrix: 1) "Is relevant information produced?" 2) "Can user benefit from research?" [66]. We could use these two questions as a conceptual tool to evaluate current research topics and propose new research strategies for Ebola. For example, sophisticated immunotherapies to treat and control Ebola can be very relevant research topics but they could be too expensive and difficult for mass production. So, some putative users could be "marginalized". If this were the case, two research strategies should be raised: the first would be aimed to improve the accessibility of immunotherapies while the second one would aimed to develop more accessible and suitable technological alternatives for the neglected stakeholders.

Matthew L. Wallace and Ismael Rafols propose that global maps of science -a scientometrics analysis tool- could "provide a sense of the range of existing theories and methodologies" (supply of knowledge) "with a connection to a given set of outcomes" (demand of knowledge) "which is conducive to identifying potential gaps and positive interactions" [67]. We consider that a combination of the methodologies used in the present paper with the use of global map of science could be a powerful strategy for analysis and management of the relation between the supply and demand of knowledge of multiple stakeholders.

### Conclusions

For the first time we have mapped the hegemonic structure of basic, clinical and patented knowledge on Ebola research. Our results suggested that Ebola research is organized around a reductionist paradigm. Three viral proteins, particularly the Ebola virus glycoprotein, and their interaction with the host immune system cells are at the core of the explanation of the disease. The involvement of the US Army is a important feature of Ebola research. The US Army is the main assignee of anti-Ebola inventions and the leading researchers are or have been related to this military institution. The lethality of Ebola and its condition as a neglected disease could be the main influence behind the reductionism and militarization of Ebola research. The knowledge structure of Ebola may be modified in the future by two fundamental events: The emergence of approved anti-Ebola treatments and the current outbreak (2013–2015) booting the clinical and public health research fields. Further research on the putative changes in the knowledge structure of Ebola would be relevant.

**Competing interests**
The authors declare that they have no competing interests.

**Authors' contributions**
All authors contributed to the interpretation of results and writing of the paper. DF-O and VMC designed the research. DF-O built the database, and performed the network analyses and text mining. All authors read and approved the final manuscript.

**Acknowledgements**
David Fajardo-Ortiz is supported by a CONACYT PhD scholarship. The Digital Medical Library of the Faculty of Medicine, UNAM provided access to the Web of Science, which enabled this research.

**Author details**
[1]Graduate program in Medical Sciences and Health, Universidad Nacional Autónoma de México, Mexico City, Mexico. [2]SAGARPA, Mexico City, Mexico. [3]Centro de Fisica Aplicada y Tecnologia Avanzada, Universidad Nacional Autonoma de Mexico, Queretaro, Mexico.

### References
1. Barry M, Traoré FA, Sako FB, Kpamy DO, Bah EI, Poncin M, et al. Ebola outbreak in Conakry, Guinea: epidemiological, clinical, and outcome features. Med Mal Infect. 2014;44(11–12):491–4. doi:10.1016/j.medmal.2014.09.009. Epub 2014 Oct 23.
2. Shears P, O'Dempsey TJ. Ebola virus disease in Africa: epidemiology and nosocomial transmission. J Hosp Infect. 2015. doi:10.1016/j.jhin.2015.01.002. [Epub ahead of print].
3. Ebola Response Team WHO. Ebola virus disease in West Africa–the first 9 months of the epidemic and forward projections. N Engl J Med. 2014;371(16):1481–95. doi:10.1056/NEJMoa1411100. Epub 2014 Sep 22.
4. Hotez P, Ottesen E, Fenwick A, Molyneux D. The neglected tropical diseases: the ancient afflictions of stigma and poverty and the prospects for their control and elimination. Adv Exp Med Biol. 2006;582:23–33.
5. MacNeil A, Rollin PE. Ebola and Marburg hemorrhagic fevers: neglected tropical diseases? PLoS Negl Trop Dis. 2012;6(6), e1546. doi:10.1371/journal.pntd.0001546. Epub 2012 Jun 26.
6. Sanchez A, Trappier SG, Mahy BW, Peters CJ, Nichol ST. The virion glycoproteins of Ebola viruses are encoded in two reading frames and are expressed through transcriptional editing. Proc Natl Acad Sci U S A. 1996;93(8):3602–7.
7. Baron RC, McCormick JB, Zubeir OA. Ebola virus disease in southern Sudan: hospital dissemination and intrafamilial spread. Bull World Health Organ. 1983;61(6):997–1003.
8. Hoenen T, Feldmann H. Ebolavirus in West Africa, and the use of experimental therapies or vaccines. BMC Biol. 2014;12(1):80 [Epub ahead of print].
9. Gupta R. Rethinking the development of Ebola treatments. Lancet Glob Health. 2014;2(10):e563-4. doi:10.1016/S2214-109X(14)70304-3.
10. Acharya M. Ebola viral disease outbreak-2014: implications and pitfalls. Front Public Health. 2014;2:263. doi:10.3389/fpubh.2014.00263. ECollection 2014.
11. Haynes DM. Still the heart of darkness: the ebola virus and the meta-narrative of disease in the hot zone. J Med Hum. 2002;23(2):133–45.
12. Murdocca C. When Ebola came to Canada: Race and the making of the respectable body. Atlantis. 2003;27(2):24–31.

13. Colhoun D. How xenophobia is driving the Ebola narrative. Columbia J Rev. 2015. Available at http://www.cjr.org/behind_the_news/racialized_ebola_narrative.php?page=all.
14. Preston R. The hot zone (1st ed.). New York: Random House; 2004.
15. Elbe S, Roemer-Mahler A, Long C. Medical countermeasures for national security: A new government role in the pharmaceuticalization of society. Soc Sci Med. 2014. doi:10.1016/j.socscimed.2014.04.035. [Epub ahead of print]
16. Gostin LO, Waxman HA, Foege W. The president's national security agenda: curtailing Ebola, safeguarding the future. JAMA. 2015;313(1):27–8. doi:10.1001/jama.2014.16572.
17. United States National Intelligence Council. National intelligence estimate: the global infectious disease threat and its implications for the United States. Environ Change Secur Proj Rep. 2000;6(1):33–65.
18. Leffel EK, Reed DS. Marburg and Ebola viruses as aerosol threats. Biosecur Bioterror. 2004;2(3):186–91.
19. Borio L, Inglesby T, Peters CJ, Schmaljohn AL, Hughes JM, Jahrling PB, et al. Working Group on Civilian Biodefense. Hemorrhagic fever viruses as biological weapons: medical and public health management. JAMA. 2002;287(18):2391–405.
20. Letter to the Speaker of the House on emergency appropriations request (and enclosures). President Barack Obama. https://www.whitehouse.gov/the-pressoffice/2014/11/05/letter-president-emergency-appropriations-request-ebola-fiscal-year-2015.
21. Shibata N, Kajikawa Y, Takeda Y, Sakata I, Matsushima K. Detecting emerging research fronts in regenerative medicine by the citation network analysis of scientific publications. Technol Forecast Soc Change. 2011;78(2):274–82.
22. Murmann JP, Frenken K. Toward a systematic framework for research on dominant designs, technological innovations, and industrial change. Res Policy. 2006;35(7):925–52.
23. Fajardo-Ortiz D, Duran L, Moreno L, Ochoa H, Castaño VM. Liposomes versus metallic nanostructures: differences in the process of knowledge translation in cancer. Int J Nanomedicine. 2014;9:2627–34. doi:10.2147/IJN.S62315.
24. Fajardo-Ortiz D, Duran L, Moreno L, Ochoa H, Castaño VM. Mapping knowledge translation and innovation processes in Cancer Drug Development: the case of liposomal doxorubicin. J Transl Med. 2014;12:227. doi:10.1186/s12967-014-0227-9.
25. Fajardo-Ortiz D, Castaño VM. Patenting Networking and Knowledge Translation in Liposomes for Cancer Therapy. Recent Patents Nanomed. 2014 [Epub ahead of print]. doi:10.2174/1877912305666150121001330
26. Falagas ME, Pitsouni EI, Malietzis GA, Pappas G. Comparison of PubMed, Scopus, Web of Science, and Google Scholar: strengths and weaknesses. FASEB J. 2008;22(2):338–42.
27. Newman ME. Power laws, Pareto distributions and Zipf's law. Contemp Phys. 2005;46:323–51.
28. Garfield E. From the science of science to Scientometrics visualizing the history of science with HistCite software. J Informetr. 2009;3:173–9.
29. Cline MS, Smoot M, Cerami E, Kuchinsky A, Landys N, Workman C, et al. Integration of biological networks and gene expression data using Cytoscape. Nat Protoc. 2007;2:2366–82.
30. Spinelli L, Gambette P, Chapple CE, Robisson B, Baudot A, Garreta H, et al. Clust&see: a Cytoscape plugin for the identification, visualization and manipulation of network clusters. Biosystems. 2013;113:91–5.
31. Newman ME. Fast algorithm for detecting community structure in networks. Phys Rev E. 2004;69:066133.
32. Doms A, Schroeder M. GoPubMed: exploring PubMed with the Gene Ontology. Nucleic Acids Res. 2005;33:W783–6.
33. Gene Ontology Consortium. The Gene Ontology project in 2008. Nucleic Acids Res. 2008;36:D440–4.
34. Lowe HJ, Barnett GO. Understanding and using the medical subject headings (MeSH) vocabulary to perform literature searches. JAMA. 1994;271:1103–8.
35. Higuchi K. Analysis of free comments in a questionnaire survey: Quantitative analysis by KH Coder. Shakai Chosa. 2012;8:92–6 (In Japanese).
36. Hachiken H, Mastuoka A, Murai A, Kinoshita S, Takada M. Quantitative analyses by text mining of journal articles on medical pharmacy. Japanese J Drug Inform. 2012;13:152–9 (In Japanese).
37. Shineha, Ryuma, Aiko Hibino, and Kazuto Kato. "Analysis of Japanese Newspaper Articles on Genetic Modification". J Sci Commun 2008, 7(2). http://jcom.sissa.it/archive/07/02/Jcom0702(2008)A02/.
38. Thomson Reuters. Derwent Innovations Index®. Available at: http://sub3.webofknowledge.com.
39. Bray M1, Davis K, Geisbert T, Schmaljohn C, Huggins J. A mouse model for evaluation of prophylaxis and therapy of Ebola hemorrhagic fever. J Infect Dis. 1999;179 Suppl 1:S248–58.
40. Xu L, Sanchez A, Yang Z, Zaki SR, Nabel EG, Nichol ST, et al. Immunization for Ebola virus infection. Nat Med. 1998;4(1):37–42.
41. Department of Microbiology and Immunology. Thomas W Geisbert profile available at http://microbiology.utmb.edu/faculty/Geisbert.asp.
42. Geisbert TW, Jahrling PB, Hanes MA, Zack PM. Association of Ebola-related Reston virus particles and antigen with tissue lesions of monkeys imported to the United States. J Comp Pathol. 1992;106(2):137–52.
43. Le Guenno B, Formenty P, Wyers M, Gounon P, Walker F, Boesch C. Isolation and partial characterisation of a new strain of Ebola virus. Lancet. 1995;345(8960):1271–4.
44. Timmins J, Scianimanico S, Schoehn G, Weissenhorn W. Vesicular release of ebola virus matrix protein VP40. Virology. 2001;283(1):1–6.
45. Jasenosky LD, Neumann G, Lukashevich I, Kawaoka Y. Ebola virus VP40-induced particle formation and association with the lipid bilayer. J Virol. 2001;75(11):5205–14.
46. Basler CF1, Wang X, Mühlberger E, Volchkov V, Paragas J, Klenk HD, et al. The Ebola virus VP35 protein functions as a type I IFN antagonist. Proc Natl Acad Sci U S A. 2000;97(22):12289–94.
47. Geisbert TW, Hensley LE, Larsen T, Young HA, Reed DS, Geisbert JB, et al. Pathogenesis of Ebola hemorrhagic fever in cynomolgus macaques: evidence that dendritic cells are early and sustained targets of infection. Am J Pathol. 2003;163(6):2347–70.
48. Mahanty S, Hutchinson K, Agarwal S, McRae M, Rollin PE, Pulendran B. Cutting edge: impairment of dendritic cells and adaptive immunity by Ebola and Lassa viruses. J Immunol. 2003;170(6):2797–801.
49. Basler CF, Mikulasova A, Martinez-Sobrido L, Paragas J, Mühlberger E, Bray M, et al. The Ebola virus VP35 protein inhibits activation of interferon regulatory factor 3. J Virol. 2003;77(14):7945–56.
50. Baize S, Leroy EM, Georges-Courbot MC, Capron M, Lansoud-Soukate J, Debré P, et al. Defective humoral responses and extensive intravascular apoptosis are associated with fatal outcome in Ebola virus-infected patients. Nat Med. 1999;5(4):423–6.
51. Geisbert TW, Pushko P, Anderson K, Smith J, Davis KJ, Jahrling PB. Evaluation in nonhuman primates of vaccines against Ebola virus. Emerg Infect Dis. 2002;8(5):503–7.
52. Yang ZY, Duckers HJ, Sullivan NJ, Sanchez A, Nabel EG, Nabel GJ. Identification of the Ebola virus glycoprotein as the main viral determinant of vascular cell cytotoxicity and injury. Nat Med. 2000;6(8):886–9.
53. Martin JE, Sullivan NJ, Enama ME, Gordon IJ, Roederer M, Koup RA, et al. A DNA vaccine for Ebola virus is safe and immunogenic in a phase I clinical trial. Clin Vaccine Immunol. 2006;13(11):1267–77. Epub 2006 Sep 20.
54. Fajardo-Ortiz D, Ochoa H, García L, Castaño V. [Translation of knowledge on cervical cancer: is there a gap between research on causes and research on patient care?]. Cad Saude Publica. 2014;30(2):415–26. doi:10.1590/0102-311X00168512. (In Spanish).
55. Merlin T, Weston A, Tooher R. Extending an evidence hierarchy to include topics other than treatment: revising the Australian 'levels of evidence'. BMC Med Res Methodol. 2009;9:34. doi:10.1186/1471-2288-9-34.
56. Cohen J, Kupferschmidt K. Infectious Diseases. Ebola vaccine trials raise ethical issues Science. 2014;346(6207):289–90. doi:10.1126/science.346.6207.289.
57. Rid A, Emanuel EJ. Ethical considerations of experimental interventions in the Ebola outbreak. Lancet. 2014;384(9957):1896–9. doi: 10.1016/S0140-6736(14)61315-5. Epub 2014 Aug 22.
58. Hantel A, Olopade CO. Drug and vaccine access in the Ebola epidemic: advising caution in compassionate use. Ann Intern Med. 2015;162(2):141–2. doi:10.7326/M14-2002.
59. Edgar EP. USAMRIID: Biodefense Solutions to Protect our Nation. Wellcome letter to the website of the U.S. Army Medical Research Institute of Infectious Diseases (USAMRIID). Available at: http://www.usamriid.army.mil/.
60. Takada A, Robison C, Goto H, Sanchez A, Murti KG, Whitt MA, et al. A system for functional analysis of Ebola virus glycoprotein. Proc Natl Acad Sci U S A. 1997;94(26):14764–9.
61. Simmons G, Wool-Lewis RJ, Baribaud F, Netter RC, Bates P. Ebola virus glycoproteins induce global surface protein down-modulation and loss of cell adherence. J Virol. 2002;76(5):2518–28.

62. Warfield KL, Bosio CM, Welcher BC, Deal EM, Mohamadzadeh M, Schmaljohn A, et al. Ebola virus-like particles protect from lethal Ebola virus infection. Proc Natl Acad Sci U S A. 2003;100(26):15889–94.
63. Bente D, Gren J, Strong JE, Feldmann H. Disease modeling for Ebola and Marburg viruses. Dis Model Mech. 2009;2(1–2):12–7. doi:10.1242/dmm.000471.
64. Feldmann H, Geisbert TW. Ebola haemorrhagic fever. Lancet. 2011;377(9768):849–62. doi:10.1016/S0140-6736(10)60667-8.
65. Strauss S. Ebola research fueled by bioterrorism threat. CMAJ. 2014;186(16):1206. doi:10.1503/cmaj.109-4910. Epub 2014 Oct 6.
66. Sarewitza D, Pielke RAJ. The neglected heart of science policy: reconciling supply of and demand for science. Environ Sci Policy. 2007;10(1):5–16.
67. Wallace ML, Rafols I. "Research Portfolios in Science Policy: Moving from Financial Returns to Societal Benefits". 2014. Available at SSRN: http://dx.doi.org/10.2139/ssrn.2500396.
68. Saito R, Smoot ME, Ono K, Ruscheinski J, Wang PL, Lotia S, et al. A travel guide to cytoscape plugins. Nat Methods. 2012;9(11):1069–76.